\documentclass[iop]{emulateapj}
\usepackage{epsfig}
\usepackage{url}
\journalinfo{accepted version}
\submitted{accepted by ApJ Letters}
\newcommand{\psr}{PSR~B1259$-$63}
\newcommand{\sta}{LS~2883}
\newcommand{\binary}{PSR~B1259$-$63/LS~2883}
\newcommand{\gr}{$\gamma$-ray}
\newcommand{\grs}{$\gamma$-rays}

\begin{document}

\title{Discovery of GeV \gr~emission from \psr/\sta}

\author{P.~H.~T. Tam$^{1}$, R.~H.~H. Huang$^{1}$, J. Takata$^{2}$, C.~Y. Hui$^{3}$, A.~K.~H. Kong$^{1,4}$, and K.~S. Cheng$^{2}$}
\affil {$^1$ Institute of Astronomy and Department of Physics, National Tsing Hua University, Hsinchu, Taiwan\\
$^2$ Department of Physics, University of Hong Kong, Pokfulam Road, Hong Kong\\
$^3$ Department of Astronomy and Space Science, Chungnam National University, Daejeon, Republic of Korea\\
$^4$ Golden Jade Fellow of Kenda Foundation, Taiwan
}
\email{phtam@phys.nthu.edu.tw}

\begin{abstract}
The binary system \binary~consists of a 47.8~ms radio pulsar that orbits the companion Be star with a period of
3.4 years in a highly eccentric orbit. The system has been well sampled in radio, X-ray, and TeV $\gamma$-ray bands, and shows orbital phase-dependent variability in all observed frequencies. Here we report on the discovery of $>$100~MeV $\gamma$-rays from \binary~through the 2010 periastron passage. Using data collected with the Large Area Telescope aboard Fermi from 33 days before periastron to 75 days after periastron, \binary~was detected at a significance of 13.6 standard deviations. The $\gamma$-ray light curve was highly variable over the above period, with changing photon index that correlates with the \gr~flux. In particular, two major flares that occur after the periastron passage were observed. The onset of \gr~emission occurs close to, but not at the same orbital phases as, the two disk passages that occur $\sim$1~month before and $\sim$1~month after the periastron passage. The fact that the GeV orbital light curve is different from that of the X-ray and TeV light curves strongly suggests that GeV \gr~emission originates from a different component. We speculate that the observed GeV flares may be resulting from Doppler boosting effects.

\end{abstract}


\section{Introduction}

The binary system \binary~comprises a young radio pulsar with a period 47.8~ms and a Be star \sta. With an eccentric (e$\sim$0.87) orbit, the pulsar approaches the periastron every 3.4 years~\citep{1259_radio_94}. The high spin-down power of \psr~\citep[$\sim$8$\times$10$^{35}$~erg~s$^{-1}$;][]{Man95} suffices to generate a relativistic pulsar wind (PW). The system is highly variable over an orbital period in radio~\citep[e.g.,][]{1259_radio_05}, X-rays~\citep[e.g.,][]{Hirayama99,Chernyakova_06,Chernyakova_09}, and TeV $\gamma$-rays~\citep{hess_1259_05,hess_1259_09}. The broadband electromagnetic spectrum is believed to result from the interaction of the PW and the stellar wind of \sta, the latter being composed of a polar wind and a dense equatorial circumstellar disk. The stellar disk is inclined with respect to the orbital plane~\citep{wex_tilted_98,1259_radio_99} such that the pulsar passes through the disk shortly before and shortly after the periastron passage.

The X-ray flux changes significantly with orbital phase. The 1--10~keV flux increases from $\sim$10$^{-12}~\mathrm{erg}~\mathrm{cm}^{-2}~\mathrm{s}^{-1}$ at apastron to more than $\sim$10$^{-11}~\mathrm{erg}~\mathrm{cm}^{-2}~\mathrm{s}^{-1}$ shortly before and shortly after the periastron passage~\citep[][and references therein]{Chernyakova_09}. The X-ray photon index ($\Gamma_X$) also shows variability related to the orbital period and a spectral state with the hardest spectra ($\Gamma_X\sim1.2$) occurred around the same time as the observed rapid growth of the X-ray flux~\citep{Chernyakova_06}. The unpulsed radio flux increases through the past four periastron passages with respect to other orbital phases, although the light curves show slightly different behaviors during each passage~\citep[][and references therein]{1259_radio_05}.

The first detection of \binary~in $\gamma$-rays was made by the H.E.S.S. Cherenkov array through the 2004 periastron passage~\citep{hess_1259_05}, and subsequently in 2007~\citep{hess_1259_09}. The 2004 and 2007 data show that the TeV emission peaks $\sim$10 days before and $\sim$20 days after the periastron passages and the possible dips seen in the TeV light curves before and after the periastron seem to coincide with the stellar disk passage~\cite[see Fig.~1 in][as well as Fig.~\ref{main_plot}]{Kerschhaggl_11}.

An upper limit of $\sim$10$^{-10}$~erg~cm$^{-2}$~s$^{-1}$ was derived using EGRET data~\citep{3rd_egret_cat} for the average $\gamma$-ray emission. A claim made by the AGILE collaboration on the detection of \binary~above 100~MeV in a 2-day period in August 2010~\citep{AGILE_1259} has not been confirmed by data collected using the more sensitive Fermi/LAT detector during the same period~\citep{LAT_1259_August}.

During a regular monitoring of $\gamma$-ray emission from \psr~shortly before the periastron passage in mid-December, we found the first evidence ($\sim$4$\sigma$) for $\gamma$-ray emission from \binary~during a 3-day time interval~\citep{Tam_1259_atel}. The discovery was later confirmed by \citet{Abdo_2nd_tel} who used data collected over 30 days from 2010 November 18.

In this Letter, detailed Fermi analysis results of \binary~from 2008 through early 2011 are presented.

\section{Fermi/LAT observations and results}

The Large Area Telescope (LAT) aboard the Fermi Gamma-ray Space Telescope can detect $\gamma$-rays with energies between $\sim$20~MeV and $>$300~GeV~\citep{lat_technical}. The $\gamma$-ray data used in this work were obtained between 2008 August 4 and 2011 February 28. These data are available at the Fermi Science Support Center\footnote{\url{http://fermi.gsfc.nasa.gov/ssc/}}. Due to the discovery of \grs~from \binary, a modified sky survey was commenced from 2010 December 27 for 10 days. In this mode the southern hemisphere receives 30\% extra exposure. We used the Fermi Science Tools v9r18p6 package to reduce and analyze the data in the vicinity of \binary. Only events that are classified as the ``diffuse'' class or the ``data-clean'' class were used. To reduce the contamination from Earth albedo $\gamma$-rays, we excluded events with zenith angles greater than 105$^\circ$. The instrument response functions ``P6\_V3\_DIFFUSE'' were used.

We carried out unbinned maximum-likelihood analyzes (\emph{gtlike}) of the circular region with a 15$^\circ$ radius centered on the \gr~position of \psr~(see below). We subtracted the background contribution by including the Galactic diffuse model (gll\_iem\_v02.fit) and the isotropic background (isotropic\_iem\_v02.txt), as well as all sources in the first Fermi/LAT catalog~\citep[1FGL;][]{lat_1st_cat} within the circular region of 25$^\circ$ radius centered on the \gr~position of \psr. We assumed single power laws for all 1FGL sources considered, except for $\gamma$-ray pulsars of which the spectra follow power laws with exponential cut-off~\citep{lat_1st_psr_cat}. The spectral parameter values for sources within 10$^\circ$ from \psr~as well as the normalization parameters of the diffuse components were set free.

No significant emission was found before 2010 November 11. We derived upper limits of $\gamma$-ray flux from \binary~using data obtained between 2008 August 4 and 2010 November 10 (see Fig.~\ref{main_plot}). However, the source became active roughly when \psr~entered the stellar disk~\citep{Tam_1259_atel,Abdo_2nd_tel}.

We first analyzed 0.2--100~GeV data from 2010 November 11 to 2011 February 28. Using a power-law description of \psr, the maximized \emph{test-statistic} (TS) value~\citep{Mattox_96} obtained for the pulsar position is 184, corresponding to a detection significance of 13.6$\sigma$. This allows us to classify \binary~as a new GeV \gr~source. The best-fit position of the $\gamma$-ray emission is estimated by \emph{gtfindsrc} to be at right ascension (J2000) $=$ 195$\fdg$67 and declination (J2000) $=$ $-$63$\fdg$73 with statistical uncertainty 0$\fdg$06 (0$\fdg$14) at the 68\%(95\%) confidence level, which is consistent with the position of \psr. The systematic uncertainty is estimated to be up to $\sim$40\%~\citep{bsl_lat}.

\subsection{Light curve}

\begin{figure}
\centerline{
\epsfig{figure=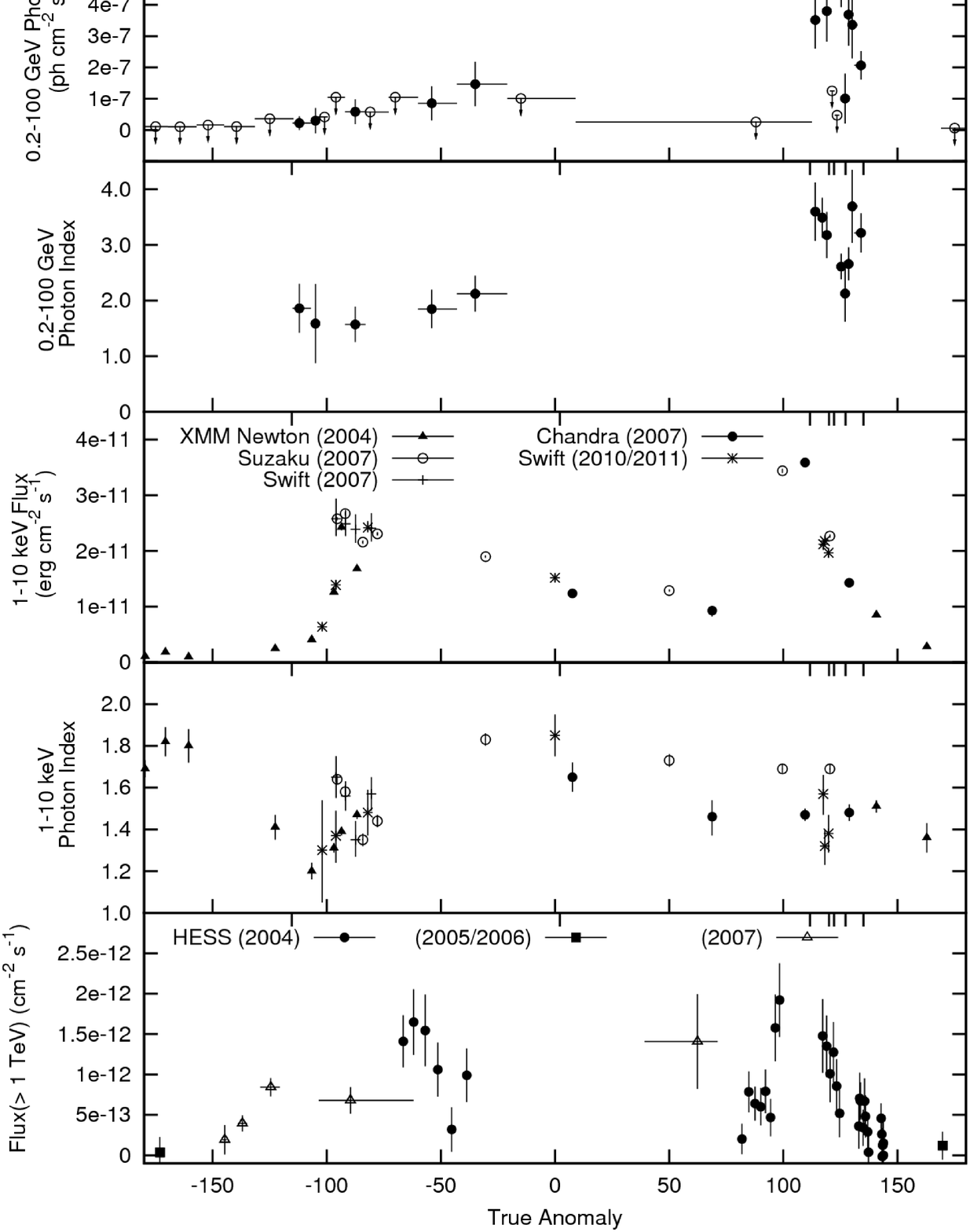,width=10cm}}
      \caption{\footnotesize X-ray, GeV and TeV observations of \binary. The error bars represent 68\% confidence-level uncertainties. (a) Light curve derived from Fermi/LAT data taken from 2008 August 4 to 2011 February 21. For better visualization in the true anomaly scale, bins span variable time duration: between 2010 November 18 and 2011 February 9, three-day bins were used, apart from the quiescent period (Q1) between 2010 December 15 and 2011 January 13.
      Power law is assumed in deriving the flux with the photon index being free. For periods without significant detection (i.e., TS$<$5), 2 confidence-level upper limits were calculated assuming $\Gamma_\gamma=2.1$; (b) LAT photon index variation over time; (c) X-ray light curve over the 2010 and previous periastron passages. 2004 XMM-Newton observations (Chernyakova et al. 2006) are marked with triangles, 2007 Suzaku observations with circles, 2007 Swift observations with crosses, 2007 Chandra observations with solid circles (Chernyakova 2009), and 2010/2011 Swift observations with stars (this work); (d) Evolution of the X-ray photon index. Notations are the same as in panel (c); (e) Gamma-ray light curve above 1~TeV. The measurements were carried out by H.E.S.S. through the 2004~\citep[fluxes using a bin width of 2 days;][]{hess_1259_05} and 2007 (monthly fluxes) periastron passages~\citep{hess_1259_09}. Also shown are upper limits derived from 2005 and 2006 observations~\citep{hess_1259_09}.}
      \label{main_plot}
   \end{figure}

We then derived a 0.2--100~GeV light curve composed mostly of 3-day bins (Fig.~\ref{main_plot}). Such bin size ensures enough photon statistics in each bin without loosing information on short time-scale variability. Finer binning (e.g. 1-day) results in low significance for most of the days and high statistical uncertainties in deriving fluxes and photon indices, except for the flaring period that occur between 2011 January 14 and February 3. For data taken more than one month before and two months after the 2010 periastron passage, larger bins were employed for better visualization. From 2010 December 15 to 2011 January 13, no evidence for \gr~emission was found, as first noted in~\citet{Kong_1259_atel}. We therefore derive an upper limit for the above quiescent period. Data points represent the flux values with TS$>$5, for which photon indices are also shown. Otherwise, an upper limit is derived.

\begin{figure}
\centerline{
\epsfig{figure=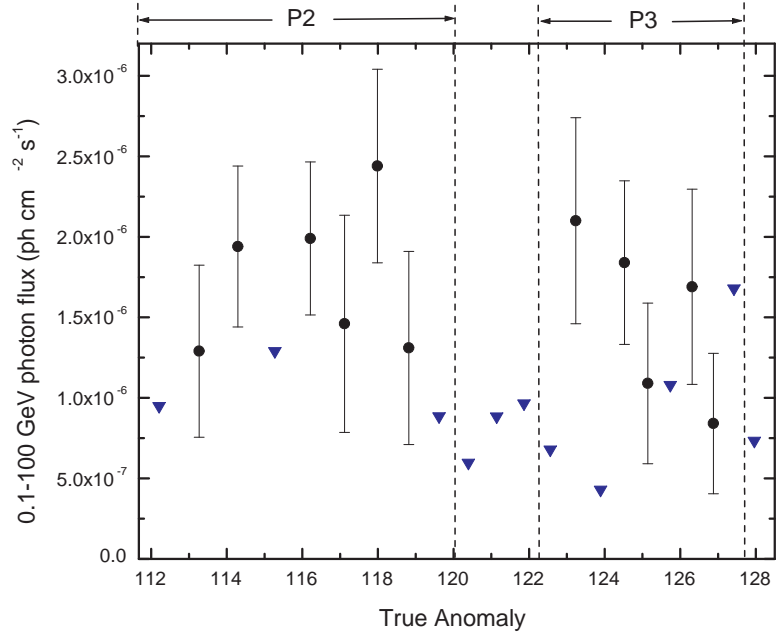,width=12cm}}
      \caption{Daily light curve from January 14 to February 4, covering flaring periods P2 and P3. For those days without significant detection (i.e., TS$<$5), two confidence-level upper limits were calculated assuming $\Gamma_\gamma=2.8$.}
      \label{zoomin_plot}
   \end{figure}

As shown in Fig.~\ref{main_plot}, the $\gamma$-ray light curve is highly variable through the 2010 periastron passage: (1) The source started to be active in \grs~about a month before 2010 periastron passage (P1); (2) It remains undetected for about one month since mid-December (Q1); (3) Subsequently, a major flaring period was identified; it peaks at $\sim$35 days after periastron. Having an average flux higher than that in P1 by an order of magnitude, this flare lasted for only $\sim$7 days (P2); (4) A second flare that peaks at $\sim$46 days after periastron, however, lasted longer, so that the source was detected until end of February (P3 and P4). To probe shorter time-scale variability during the flaring period (i.e., P2 and P3), we also produced the light curve with 1-day binning during January~14 to February~4 (Fig.~\ref{zoomin_plot}). Here 0.1--100~GeV photons were used to increase photon statistics. To better demonstrate the variability, we plot data points with TS$>$5 (rather than the more standard criterium of TS$>$9) as flux values since a substantial number of data are of TS between 5 and 9. It can be seen that \grs~from \binary~undergo rapid variations down to time scale of one day. Such variability is one of the fastest detected from any Galactic GeV source on the sky. No emission was detected during January 22--26, indicating that the source underwent a short quiescent period between the two major flares.

We also show in Fig.~\ref{main_plot} the orbital X-ray and TeV light curves. The GeV orbital light curve is different than the X-ray and TeV light curves, suggesting that the origin of GeV \grs~is different than the others.

\subsection{Spectral analysis}
As the derived photon indices change significantly with time, we defined four periods that are indicated in Figure~\ref{main_plot} and Table~\ref{4P}. We further divided the 0.1-300~GeV $\gamma$-rays arriving during P1, P2, P3, and P4, respectively, into six energy bins of logarithmically equal bandwidths and reconstructed the flux using \emph{gtlike} for each band independently. A power-law (PL) model for each bin was assumed and the photon spectral index was fixed at a representative value $\Gamma_\gamma=$2.8. While the emission is detected (i.e., TS$>$5) from 1.4~GeV to 20~GeV for P1, no $\gamma$-ray source was apparent at the \psr~position in the four bins above 1.4~GeV (the derived TS values $<$5) in the likelihood analysis for P2, P3, and P4, indicating a cut-off at energy $\sim$1~GeV during the flaring period. See Fig.~\ref{SED} for the spectrum derived from a combined analysis of P2 and P3. We therefore attempted to fit the 0.1--100~GeV spectrum with a PL with an exponential cut-off (PLE), as well as a broken PL. We found that PLE describes the spectrum even better during the periods P2, P3, and P4, by $\Delta$TS$=$TS$_\mathrm{PLE}-$TS$_\mathrm{PL}\ge$8, i.e., $\ga$3$\sigma$ in significance levels. The best-fit parameters are shown in Table~\ref{4P}. In particular, the cut-off energies were found to be 310$\pm$160~MeV, 550$\pm$330~MeV, and 250$\pm$95~MeV during the periods P2, P3, and P4, respectively. On the other hand, the parameters provided by the broken PL with all four parameters being free are not well constrained; we therefore do not consider this model.

At a distance 2.3~kpc~\citep{Neg11}, the average energy flux during the flares of $\sim$3$\times10^{-10}$~erg~cm$^{-2}$~s$^{-1}$ corresponds to the \gr~luminosity 1.9$\times10^{35}$~erg~s$^{-1}$. Given the pulsar spin-down luminosity $\sim$8$\times$10$^{35}$~erg~s$^{-1}$, the average \gr~efficiency is about 25\% during the flares.

To demonstrate that GeV emission originates from a component that is temporally and spectrally different than the TeV emission, we put together the 100~MeV to 300~GeV spectra and the $>$300~GeV \gr~spectra obtained for different orbital phases in Figure~\ref{SED}. It is clear that the GeV emission evolves differently compared to the TeV emission, assuming that the TeV behavior does not change dramatically between 2004 and 2010 periastron passages.

\subsection{Correlation between flux and photon index}
In Fig.~\ref{correlation_plot} we plot \gr~flux versus photon index, showing a correlation between these two quantities. We carried out a nonparametric correlation analysis. The computed Spearman rank correlation coefficient between two quantities is $-0.7363$. The probability that this coefficient is different than zero is 0.9959. We have also calculated the linear correlation coefficient (i.e., Pearson's r).
This results in Pearson's r$=-0.7874$ and the probability that this coefficient is different than zero is 0.9986.

\section{Discussion}
High-energy emissions from $\gamma$-ray
binaries (PSR~B1259-63, LS~5039, and LS~I+$61^{\circ}303$) have been discussed using
leptonic models (e.g., Tavani \& Arons, 1997; Dubus 2006; Takata \& Taam 2009)
 and hadronic models (e.g., Kawachi et al. 2004; Chernyakova et al. 2009).
In leptonic models, PW particles (electrons and positrons)
 are accelerated at the shock where the dynamical
pressure of the PW and that of the stellar wind are
 in balance. These particles in turn emit non-thermal photons over a wide range
of energies via synchrotron radiation (radio to GeV \grs) and
inverse-Compton (IC) upscattering off star light
($>$10~GeV \grs), resulting in two peaks in the broadband spectrum.
In hadronic models~\citep[e.g.,][]{Chernyakova_06,Chernyakova_09}, the inverse Compton emission of high-energy electrons resulting from $\pi^0$-decay is responsible for emission from optical up to TeV energies. In this case, the broadband spectrum is rather flat in the energy range of 0.1--100~GeV.

During the pre-periastron epoch, GeV emission at a flux level $\sim6\times10^{-11}$~erg~cm$^{-2}$~s$^{-1}$ is detected from mid-November through mid-December. The fitted spectrum during this period (P1) is relatively hard ($\Gamma_\gamma\sim2$). Such hardness may be consistent with the predictions of both the hadronic and the leptonic models (see below for a discussion of the leptonic interpretation during this period).

The onset of the flare-like
 GeV emission at the true anomaly $110^{\circ}-130^{\circ}$ occurred close to
the disk passage at the true anomaly $80^{\circ}-140^{\circ}$, suggesting
that the origin of the \grs~might be related to the disk passage.
On one hand, the observed cut-off at several hundred MeV does not favor hadronic models.
On the other hand, it may be difficult to explain the
flare-like GeV emission using simple leptonic models as well.
 First, leptonic models predict a cut-off around 100~MeV for a synchrotron spectrum~\citep[e.g.,][]{Takata09}. Second, the stellar disk pushes the shock towards the pulsar when the pulsar encounters the disk. Although the increase of the magnetic field enhances
the synchrotron power, it also reduces
the synchrotron cooling time of the accelerated particles.
As GeV photons are emitted by fast-cooling particles, these two effects should compensate
each other. Consequently, the shift of the shock position due to
the stellar disk cannot enhance the GeV emission to the level that we observe.

Doppler boosting may provide a plausible mechanism
to produce the GeV flares during January~14 -- February~3, 2011~\citep[see also][]{Kong_sw_11}.
Numerical simulations in the hydrodynamic limit imply that the post
shock bulk flow for the binary system can be accelerated into relativistic
regime because of a rapid expansion of the flow
in the downstream region (Bogovalov et al. 2008). For example,
 Dubus et al. (2010) discussed the effects of Doppler
boosting on the  orbital modulations of the X-ray and TeV emission.
 If the bulk flow  is relativistic and oriented
radially away from the star,  the Doppler effect may be at work close to the true anomaly corresponding to the direction
 of the Earth ($\sim 130^{\circ}$). This is approximately the time when
the GeV flares were observed (i.e., $110^{\circ}-130^{\circ}$). The Doppler effect increases the photon energy
as $E={\cal D}E'$ and  the intensity  as $I_{\nu}\propto{\cal D}^{3+\alpha}I'_{\nu}$, where ${\cal D}$ is the Doppler factor, and
$\alpha=\Gamma-1$ the spectral index. Non-prime and prime notations represent quantities
 in the observer and co-moving frames, respectively.

In \grs, $\alpha_\gamma\sim1-2$ is expected from synchrotron radiation models, which is consistent with the Fermi results.
An enhancement factor of 5--10 in flux during flares (P2 and P3) compared to the emission before periastron (P1; see table~\ref{4P} for details) suggests ${\cal D}\sim$1.5--2.

Suzaku observations indicate a low energy break, i.e., around 10~keV~\citep{Suzaku09}.
Because the photon index of the synchrotron spectrum below the break is
$\alpha=-1/3$, the enhancement factor becomes $D^{3-1/3}\sim3$. Thus, the enhancement in X-rays is
suppressed compared to \grs. It is important to obtain simultaneous observations in X-rays and
gamma-rays during the flaring period to test the boosting model.

\begin{figure}
\centerline{
\epsfig{figure=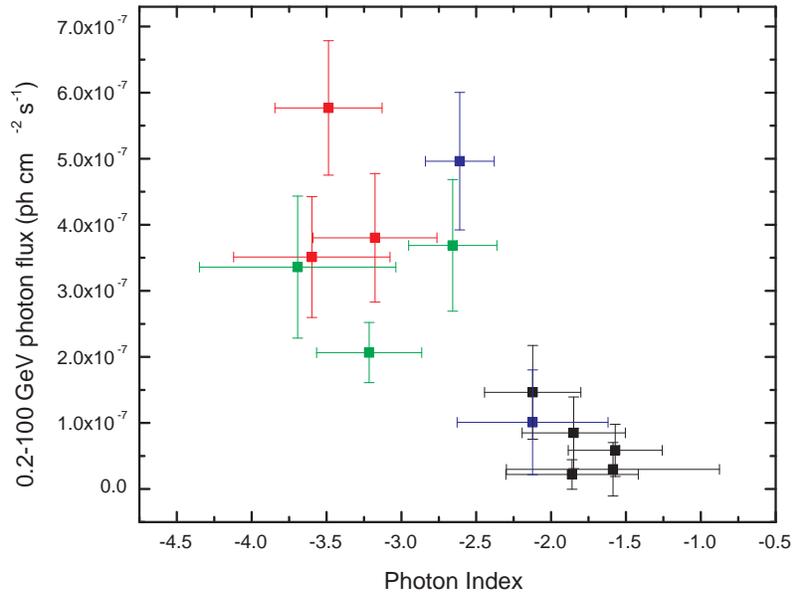,width=12cm}}
      \caption{Photon flux versus photon index in the 0.2--100~GeV band. Data are binned in the same manner as in Figure~\ref{main_plot}. Data obtained during periods P1, P2, P3, and P4 are shown in black, red, blue, and green, respectively.}
      \label{correlation_plot}
   \end{figure}

\begin{figure}
\centerline{
\epsfig{figure=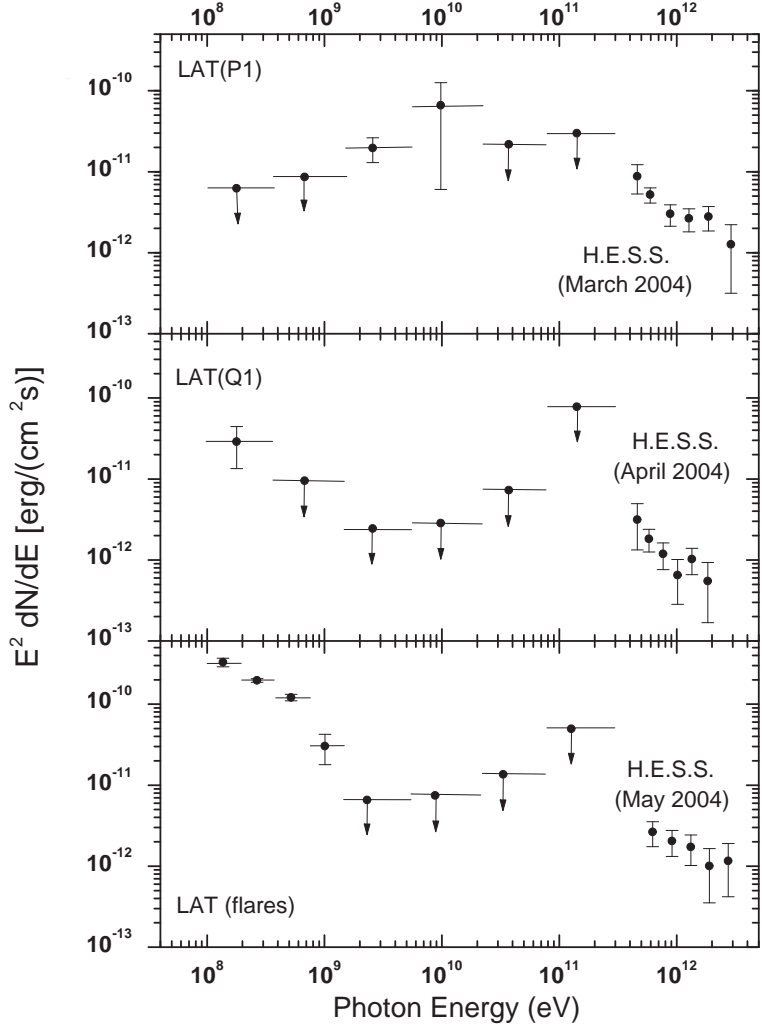,width=12cm}}
      \caption{\footnotesize 100~MeV to $\sim$3~TeV \gr~spectra in $E^2\,dN/dE$ representation during different orbital phase periods. Data between 100~MeV and 300~GeV were obtained by Fermi/LAT through the 2010 periastron passage. Data above 300~GeV were obtained by the H.E.S.S. telescopes through the 2004 periastron passage (reproduced from Fig. 4 in~\citet{hess_1259_05}). The observations were carried out in the periods shown in each panel which are chosen to be in the similar orbital phases for the two energy bands in the same panel. The LAT data were divided into six energy bins of logarithmically equal bandwidths and the flux in each bin was reconstructed using \emph{gtlike} independently, assuming power law within each bin and fixing $\Gamma_\gamma=2.1$ for P1 and Q1, and $\Gamma_\gamma=2.9$ for the flaring period (January 14 -- February 3). Two confidence-level upper limits were calculated for those bins with TS$<$5. \emph{Middle panel}: \grs~at the lowest bin 100--380~MeV are (marginally) detected at TS$=$10, i.e. significance level $\sim$3. \emph{Bottom panel}: the lowest two energy bands are further divided into two bins each to increase the spectral resolution.}
      \label{SED}
   \end{figure}

In leptonic models, the correlation between $\Gamma_\gamma$ and flux (Fig.~\ref{correlation_plot}) may  be
understood as follows. If the Doppler effect is responsible for the flares, photons are boosted to higher energies and the observed synchrotron flux is amplified. However, the Doppler boost does not affect much the photon flux of the IC component,
since more scatters between particles and
incident photons would occur in the Klein-Nishina regime.
Consequently, the  increase
in the IC flux by the Doppler boosting and the decrease due to the
Klein-Nishina effect will compensate each other. The energy flux of the  synchrotron radiation becomes much higher than the
IC radiation, causing the 0.1--100~GeV spectrum to be dominated by the high-energy tail of the synchrotron spectrum. This gives rise to the steep spectrum ($\sim$3; corresponding to flare emission) shown in Fig.~\ref{correlation_plot}.
The emission before periastron (with $\Gamma_\gamma\sim$2)
may be related to emission originating just behind the shock
where the bulk Lorentz factor is $\sim$1.
Some leptonic models (e.g. Takata et al. 2009;
Dubus et al. 2010) have predicted similar energy fluxes of synchrotron
and inverse-Compton radiation just behind the shock, hence $\Gamma_\gamma\sim2$, which is close to the observed value in the power-law fit. Although there is no evidence for two spectral components during the pre-periastron period (see Fig.~\ref{SED}), this possibility cannot be excluded given the relatively low significance detection, i.e., $\sim$5$\sigma$,
during this period.

\binary~is the third known binaries with significant detection in GeV, after LS~I+$61^{\circ}303$~\citep{lat_ls61_303} and LS~5039~\citep{lat_ls5039}. It may be intuitive to compare the three $\gamma$-ray
binaries:

\begin{enumerate}
  \item For LS~I+61 303 and PSR~B1259-63, the GeV gamma-ray peak occurs after periastron (and before apastron), though not in the same orbital phase. For LS~5039, the gamma-ray emission peaked at periastron;
  \item While no orbital phase-related spectral change has been reported in LS~I+$61^{\circ}303$, the ``softer when brighter'' behavior of GeV \grs~has been found for both LS~5039 and PSR~B1259-63;
  \item In all three systems, spectral cutoff is observed at GeV energies during at least part of the orbit;
  \item In all three systems, the X-ray and TeV light curves are more correlated whereas GeV \grs~always come out in different orbital phases~\citep{ls5039_xray_tev,ls61_303_xray_tev}. This strongly suggests that GeV \grs~originate from components different than the other two energy bands.
\end{enumerate}

\binary~is the only system among these three for which the nature of the compact object is certain. Given the similarities of some emission features of these systems, the results presented here should shed light on the GeV radiation mechanism of the other systems.

\acknowledgments
We acknowledge the use of data and software facilities
from the FSSC, managed by the HEASARC at GSFC. JK and KSC are supported by a GRF grant of HK Government under HKU700908P, and AKHK is supported partly by the National Science Council of the Republic of China (Taiwan) through grants NSC99-2112-M-007-004-MY3 and NSC100-2923-M-007-001-MY3. CYH is supported by research fund of Chungnam National University in 2010.

\begin{table}
\begin{center}
\caption{0.1--100~GeV \gr~properties during different periods through the 2010 periastron passage. \label{4P}}
\begin{tabular}{l@{ }c@{ }c@{ }c@{ }l@{}cc@{}c@{ }c@{}r@{ }r}
    \tableline\tableline
    Period & Year & Date & True Anomaly & Model & Photon Flux & Energy Flux & Photon Index & Cutoff Energy & TS/ & $\Delta$TS/ \\
           &      &     &     &   & (cm$^{-2}$~s$^{-1}$) & (erg~cm$^{-2}$~s$^{-1}$) &           & (MeV)     & sig. & sig. \\
    \tableline
    P1 & 2010 & Nov 11 -- Dec 14 & -115.3--2.1 & PL & (9.9$\pm$4.1)$\times$10$^{-8}$ & (5.8$\pm$1.6)$\times$10$^{-11}$ & 2.1$\pm$0.2 & & 31/5.6$\sigma$ & \\
    \tableline
    P2 & 2011 & Jan 14 -- Jan 22 & 111.7--120.0 & PL & (1.5$\pm$0.2)$\times$10$^{-6}$ & (2.9$\pm$0.3)$\times$10$^{-10}$ & 3.0$\pm$0.1 & & 163/12.8$\sigma$ &  \\
       &      &              &            & PLE & (1.3$\pm$0.2)$\times$10$^{-6}$ & (2.7$\pm$0.3)$\times$10$^{-10}$ & 1.8$\pm$0.6 & 310$\pm$160 & 184/13.6$\sigma$ & 21/4.6$\sigma$ \\
    \tableline
    P3 & 2011 & Jan 26 -- Feb 3 & 122.2--127.2 & PL & (1.4$\pm$0.2)$\times$10$^{-6}$ & (3.1$\pm$0.4)$\times$10$^{-10}$ & 2.8$\pm$0.1 & & 109/10.4$\sigma$ & \\
           &      &           &               & PLE & (1.3$\pm$0.2)$\times$10$^{-6}$ & (2.8$\pm$0.3)$\times$10$^{-10}$ & 2.0$\pm$0.5 & 550$\pm$330 & 117/10.8$\sigma$ & 8/2.8$\sigma$ \\
    \tableline
    P4 & 2011 & Feb 4 -- Feb 21 & 127.2--135.1 & PL & (8.7$\pm$1.1)$\times$10$^{-7}$ & (1.9$\pm$0.2)$\times$10$^{-10}$ & 2.8$\pm$0.1 & & 97/9.9$\sigma$ & \\
           &      &           &               & PLE & (7.5$\pm$1.1)$\times$10$^{-7}$ & (1.7$\pm$0.2)$\times$10$^{-10}$ & 1.1$\pm$0.6 & 250$\pm$95 & 117/10.8$\sigma$ & 20/4.5$\sigma$ \\
    \tableline
\end{tabular}
\end{center}
\end{table}

\end{document}